\documentclass[aps,pra,reprint,twocolumn,showpacs,floatfix,superscriptaddress]{revtex4-1}
\usepackage{amssymb,amsmath,amstext}                %%   American Physical Society math etc extensions
\usepackage{graphicx}                                               %%   Include figure files                                            
\usepackage{epstopdf}                                               %%   help with eps -> pdf 
\usepackage[english]{babel}
\usepackage{color}                                                     %%   change color
\usepackage{bm}                                                        %%   bold math                                    
\usepackage{appendix}                                              %%   formating appendicies
\usepackage[utf8]{inputenc}
\usepackage{bbold}
\usepackage{bbm}

\usepackage{ulem}
\usepackage{latexsym}
\usepackage{xcolor}
\usepackage{braket}
\usepackage{subfigure}
\definecolor{lblue} {RGB}{51,71,158}

\usepackage[colorlinks=true,citecolor=blue,linkcolor=blue,urlcolor=lblue]{hyperref}

\newcommand{\be}{\begin{equation}}
\newcommand{\ee}{\end{equation}}

\begin{document}

\title{Quantum chaos and level dynamics}
% Force line breaks with \\

\author{Jakub Zakrzewski}
\email{jakub.zakrzewski@uj.edu.pl}
\affiliation{Institute of Theoretical Physics, Faculty of Physics, Astronomy and Applied Computer Science, Jagiellonian University,  \L{}ojasiewicza 11, 30-348 Krak\'ow, Poland }
\affiliation{Mark Kac Complex
Systems Research Center, Jagiellonian University, 30-348 Krak\'ow,
Poland. }

\date{\today}% It is always \today, today,

%\pacs{03.67.Lx, 42.50.Dv}% PACS, the Physics and Astronomy
                              
\begin{abstract}
We review application of level dynamics to spectra of quantally chaotic systems. We show that statistical mechanics approach gives us predictions about  level statistics intermediate between integrable and chaotic dynamics. Then we discuss in detail different statistical measures involving level dynamics such as  level avoided-crossing distributions, slope and curvature of level distributions showing both the postulate of unversality and its limitations. We mention shortly the experimental confirmations of these theories. We concentrate in some detail on measures imported from quantum information approach such as the fidelity susceptibility and more generally geometric tensor matrix elements. The possible open problems are suggested. 
\end{abstract}
\maketitle

\section{Introduction}
\label{intro}

It is a great pleasure to be able to contribute to Giulio Casati 80th birthday volume. Since very beginning of my encounter with quantum chaos Giulio Casati was one of those whose works inspired younger people.
As an example let me mention a contribution of late Prot Pako\'nski with whom I had a pleasure to consider
 the  Kepler map (one of toy models of Giulio), extending it to arbitrary polarization of the microwaves \cite{Pakonski01}. In this review I  will discuss, however, a different topic - the statistical measures related to level dynamics in quantally chaotic systems. This is where we met scientifically, contributing to a single paper I had an honour to coauthor with Giulio \cite{Guarneri95}. While the subject of level dynamics developed in the 980s and 990s of the former millenium, it found interesting extensions and applications in the modern many-body physics. 

The level dynamics, described in a pedagogical way in the books of late Fritz Haake \cite{Haakebook} and Hans-J\"urgen St\"ockmann \cite{Stockmann} consider the motion of levels as a function of some arbitrary scalar parameter $\lambda$ which characterizes the Hamiltonian $H(\lambda)$ of the system under consideration. It may be viewed as 
 the motion of interacting fictitious particles (represented by levels) with   $\lambda$ being the effective time,  as described originally by Pechukas \cite{Pechukas83} and followed by  Yukawa \cite{Yukawa85} who built a corresponding statistical mechanics picture. This was applied in different ways to either justify the random matrix theory application to quantum chaotic spectra or
 define new statistical measures and find their distributions. Without any claim for completeness some of these application will be reviewed below. 
 
 We begin with defining the notation for level dynamics and the corresponding statistical mechanics
 picture in Section~\ref{statrev} showing how standard results from this approach provides a prediction for level statistics in chaotic - integrable transition. The resulting distribution seems to work surprizingly well for the data in many-body localization crossover  %  a result existing in the literature for some time \cite{Gaudin66,Forrester93,Hasegawa98} but tested recently only on numerical data
 \cite{De21}. To somehow complete the picture we review in the next Section several other models for the statistics in the transition, notably the banded matrix model, developed by Casati and coworkers \cite{Casati90,Casati91,Casati93,Casati94,Casati96}.
 In Section~\ref{sec:univ} we review the universality conjecture in level dynamics
 \cite{Simons93} showing how it is reflected in the so called curvature distributions \cite{Zakrzewski93} in Section~\ref{sec:curv}. We mention the velocity correlations in Section~\ref{sec:vel} and the avoided crossings statistics  \cite{Zakrzewski91,Zakrzewski93b} in Section~\ref{sec:avo}. Applications of these measures are discussed stressing their limitations in real systems.  Section~\ref{sec:fid} describes, on the other hand, very recent findings on distribution of fidelity susceptibility \cite{Sierant19f} while the extensions to many parameter dynamics with geometric tensor matrix elements distributions are  reviewed in Section~\ref{sec:many}.
 We mention briefly the parametric measures in the transition to localized regime
 in Section~\ref{sec:loc}.  We finalize with conclusions discussing future perspectives.

\section{Level dynamics}
\label{statrev}

Let us recall some basic details on level dynamics, to some extent to fix the notation. Let the Hamiltonian, $H(\lambda)=H_0+\lambda V$, depend on some parameter $\lambda$ for arbitrary
$H_0$ and $V$.  
The eigenvalue equation
\begin{align}
H(\lambda) |a(\lambda)\rangle=E(\lambda)_a|a(\lambda)\rangle,
\label{eigen}
\end{align}
(where $E(\lambda)_a$ is the eigenvalue corresponding to eigenvector 
$|a(\lambda)\rangle$)
upon  differentiation with respect to $\lambda$ gives
\begin{align}
\frac{d}{d\lambda} E_a \equiv \dot{E}_a=\langle a|V|a\rangle \equiv V_{aa}.
\label{e3}
\end{align}
Let us call $p_a \equiv \dot{E}_a$ a velocity of level $E_a$ where $\lambda$ becomes a fictituous time.A $\lambda$ derivative (denoted by a dot over the variable)  of $p_a$ yields 
\begin{align}
\dot{p}_{a}=2\sum_{b\neq a}\frac{V_{ab} V_{ba}}{E_a-E_b}= 2\sum_{b\neq a}\frac{|f_{ab}|^2}{(E_a-e_b)^3}
\label{e4}
\end{align}
where we introduced $f_{ab}=V_{ab}(E_a-E_b)$. Following the procedure one step further we find equations for $\dot{f_{ab}}$.
 \begin{equation}
\dot{f}_{ab}=\sum_{r\neq a,b}^{}f_{ar}f_{rb}\bigg[\frac{1}{(E_a-E_r)^2}-\frac{1}{(E_b-E_r)^2}\bigg].
\label{e5}
\end{equation}
One notices that no new quantities appear, the 
 set of equations \eqref{e3}-\eqref{e5} is closed. It is sometimes called as the {Pechukas-Yukawa} equations following original works \cite{Pechukas83,Yukawa85}.  This  set of nonlinear equations is integrable - as pointed out in \cite{Haakebook} this is not surprizing as a problem is equivalent to a diagonalization of the Hamiltonian matrix.
 
The eigenvalues of $H=H_0+\lambda V$ are $V$ dominated for large $\lambda$  and the dynamics becomes trivial.  Haake \cite{Haakebook} introduces a different $\lambda$ dependence (equivalent for small $\lambda$):  $H(\lambda)=\sqrt{f}(H_0+\lambda V)$ with $f=(1+\lambda^2)^{-1}$ while we shall follow the ``trigonometric choice''  \cite{Zakrzewski93,Zakrzewski93b,Stockmann}
\begin{equation}
H=H_0\cos(\lambda)+V\sin(\lambda).
\label{eq:ham}
\end{equation}
 This results in an additional harmonic binding of eigenvalues which prevents them from escaping to infinity. In effect the equations of motion become: 
\begin{align}
 \dot{E}_a=\langle a|\dot{H}|a\rangle = p_a
\label{e3p}
\end{align}
and 
\begin{align}
\dot{p}_{a}=- E_a + 2\sum_{b\neq a}\frac{|f_{ab}|^2}{(E_a-e_b)^3}
\label{e4p}
\end{align}
with $f_{ab}=\langle a | \dot H | b \rangle (E_a-E_b)$. 

Since the dynamics is integrable the appropriate statistical description should involve all possible constants of the motion. Such an approach would be a formidable task. The Yukawa simplified way is just to consider the simplest integrals of motion,
the total energy $H$ and the trace of the square of $f_{ab}$ matrix, $L=\frac{1}{2}\textrm{Tr}(F^2)$ \cite{Stockmann} with 
\begin{equation}
H=\frac{1}{2}\sum_{n}^{}(p_n^2+E_n^2)+\frac{1}{2}\sum_{n,m}^{}(\frac{|f_{nm}|^2}{(E_n-E_m)^2}.
\end{equation}
The phase-space density, according to Gibbs is: 
\begin{equation}
\rho=\frac{1}{Z}\exp(-\alpha H-\gamma Q).
\end{equation}
The reader may be surprized that we use $\alpha$ as an effective inverse temperature. The simple reason is that, in order to follow the sacred quantum chaos notation, we reserve $\beta$ for a level repulsion parameter with $\beta=1,2,4$ characterizing different universality classes of Dyson and corresponding, for gaussian ensembles to Gaussian Orthogonal, Unitary, and Symplectic Ensembles (denoted as GOE, GUE, and GSE, respectively). 
The density $\rho$ may be {explicitly} written out as 
\begin{equation}
\label{rho}
\begin{split}
\rho=\frac{1}{Z}\exp(-\alpha\bigg[\frac{1}{2}\sum_{n}^{}(p_n^2+E_n^2)+\frac{1}{2}\sum_{n,m}^{}(\frac{|f_{nm}|^2}{(E_n-E_m)^2}\bigg]\\
-\gamma \frac{1}{2}\sum_{n,m}^{}\bigg|f_{nm}\bigg|^2).
\end{split}
\end{equation}
 By integrating out the variables $p_n$ and $f_{nm}$, we can compute the joint probability distribution (JPD) of eigenvalues \cite{Stockmann}
\begin{equation}
\begin{split}
P(E_1,E_2,.....,E_n)\sim\prod_{n<m}^{}\bigg|\frac{(E_n-E_m)^2}{1+\frac{\gamma}{\alpha}(E_m-E_n)^2}\bigg|^{\beta/2}\\ exp\bigg(-\frac{\alpha}{2}\sum_{n}^{}E_n^2\bigg),
\end{split}
\label{s3e18}
\end{equation}
with $\beta=1,2,4$ corresponds to three universality classes. The  $\beta$ appears in \eqref{s3e18} as the integrated our ``angular momenta'' $f_{nm}$ structure depends on the universality class with $F$ being
orthogonal, unitary or symplectic.  
The similar ensemble was considered by Gaudin \cite{Gaudin66} as well as Forrester \cite{Forrester93} and Hasegawa and Ma \cite{Hasegawa98}.
They considered mainly two point correlation function for the unitary ensemble. We concentrate rather on the time-reversal invariant case, as most commonly met in current many-body localization studies.

Equation~\eqref{s3e18} is simplified in different limiting cases. The Possonian distribution is reached  in $\gamma/\alpha>>1$ limit.  The distribution becomes
\begin{equation}
P(E_1,E_2,.....,E_n)\sim\exp\big(-\frac{\alpha}{2}\sum_{n}^{}E_n^2\big).
\end{equation}
On the other hand, for $\gamma/\alpha\ll 1$
the distribution yields the Gaussian ensembe result. We have
\begin{equation}
P(E_1,E_2,.....,E_n)\sim\prod_{n>m}^{}\bigg|E_n-E_m\bigg|^{\beta/2} \exp\big(-\frac{\alpha}{2}\sum_{n}^{}E_n^2\big).
\end{equation}
Finally to reach (GOE) in this limit we fix $\beta=\alpha=1$ and denote $\gamma/\alpha=10^p$. The distribution \eqref{s3e18} takes the final form
\begin{equation}
\begin{split}
P(E_1,E_2,.....,E_n)\sim\prod_{n<m}^{}\bigg|\frac{(E_n-E_m)^2}{1+10^p (E_m-E_n)^2}\bigg|^{1/2}\\
exp\bigg(-\frac{1}{2}\sum_{n}^{}E_n^2\bigg)
\end{split},
\label{s3e21}
\end{equation}
where $p=\log_{10}\frac{\gamma}{\beta}$ is the single  parameter nterpolating between  GOE and Poisson limit. The first term in the \eqref{s3e21} signifies the pairwise interaction between the particles and the exponential term provides the harmonic binding of the eigenvalues.  The resulting distribution, obtained using Monte-Carlo sampling for different $p$ was shown to faithfully reproduce statistics of eigenvlaues on the transition between ergodic and many body localized situations \cite{De21}.

\section{Other interpolating ensembles} 
\label{histo}

It is interesting to review 
several interpolating statistics models proposed in the past. An early work of  
Rosenzweig and Porter  \cite{Rosenzweig60} is certainly worth mentioning. In this model the variance of off diagonal elements in random matrices is controlled by a matrix dimension dependent parameter.
Its value interpolates between  the gaussian orthogonal ensemble (GOE) value  to vanishing values for the Poissonian case. The other approach was proposed on the basis of Wigner-inspired
$2\times 2$ matrix  approach by Lenz and Haake \cite{Lenz91}. 
Yet another was an {\it ad hoc} expression known as Brody distribution \cite{Brody73}  which surprizingly well fitted low resolution experimental data. Berry and Robnik \cite{Berry84b} proposed the distribution  based on sound physical assumption of the separation between
``chaotic'' wavefunctions faithful to GOE and those localized in the regular part of the phase space. The corresponding distribution was shown to work well in the so called deep semiclassical limit \cite{Prosen98}. 
Another proposition due to  Seligman and coworkers \cite{Seligman84} assumed that the variance of off-diagonal elements $a_{ij}$should scale as $\exp\left[ - (i -j)^2/\sigma^2\right]$. For $\sigma\rightarrow 0 $ one recovers the Poisson case while $\sigma\rightarrow \infty$ becomes GOE.
Yet another well-known approach is that of Guhr  \cite{Guhr96} who used supersymmetric techniques to express the two-level correlation function in the Poisson-GOE ensemble in terms of a double integral. It is worth stressing that  another popular proposition was advocated by Giulio Casati and coworkers \cite{Casati90,Casati91}. They considered banded Gaussian random matrices as a useful tool in describing the transition, the corresponding parameter was $y=b^2/N$ with $b$ being the matrix bandwidth and $N$ its rank.  

While the (unfolded) level spacing statistics was the main object of quantum chaos studies,  in many-body localizaiton (MBL) context  Huse and Oganesyan \cite{Oganesyan07} introduced a new measure  - the {gap ratio}, defined as $r_n=min[\delta_n,\delta_{n-1}]/max[\delta_n,\delta_{n-1}]$, where $\delta_n=E_{n+1}-E_{n}$ is the energy gap between the consecutive energy levels. The dimensionless {gap-ratio}  does not require unfolding. The  MBL transition description was addressed by  Serbyn and Moore \cite{Serbyn16} who proposed two stages GOE-Poisson transition: (1) A Short Range Plasma Model (SRPM) and (2) a semi-poissonian level statistics  \cite{Bogomolny99,Bogomolny01}. Recent efforts worth mentioning are 
 a  $\beta$-Gaussian($\beta-G$) model was introduced \cite{Buijsman18}. A comparison of the performance of different models is given in  \cite{Sierant19b} while \cite{Sierant20}  proposes a more complicated, two parameter  $\beta-h$ model, where the pairwise interaction between the levels is limited to a number $h$. 
 
 Comparison of some of these models with numerics for interacting disordered spin systems modelling ergodic to MBL transition is given in \cite{De21}. The interested reader should consult \cite{De21} for details, it suffices to say here that the single parameter Yukawa-like model described above compared favorably with other single parameter models and quite faithfully reproduced the disordered spin data for MBL-ergodic crossover.

\section{Universality of parametric dynamics}
\label{sec:univ}

A simple inspection of  Eq.~\eqref{rho} shows that the velocities, $p_n$, have, in this approach a Gaussian distribution with the variance determined by the ``inverse temperature'' $\alpha$.  This is the essence of 
level dynamics universality as determined and thoroughly studied by Simons and Altschuler \cite{Simons93,Simons93a}.  The level spacings have a single scale - the mean level spacing, $\Delta$. 
The unfolding corresponds then to rescaling the energy levels $\epsilon_i=E_i/\Delta$. Level dynamics introduces a novel scale determining how fast 
the eigenvalues change with the parameter $\lambda$. The original definition \cite{Simons93} involves the velocity-velocity correlation function for unfolded levels
\begin{equation}
C(\lambda) = \left< p_n(0) p_n(\lambda) \right>/\Delta^2
\end{equation}  averaged over eigenstates $n$. Then $C(0)$ yields the second, apart from the mean level spacing, important scale.  When the levels are unfolded using mean spacing, $\Delta$, and parametric dependence is "unfolded" using $C(0)$ as \cite{Simons93}
\begin{equation} 
x=\sqrt{C(0)}\lambda
\label{uni}
\end{equation}
 the spectral properties of different systems should behave similarly. Clearly $C(0)$ in our notation is directly related, modulo unfolding,  to the ``inverse temperature'' $\alpha$ in the Gibbs ensamble.
% For convenience (to shorten the following expressions) we shall rather use
% \begin{equation} 
%x=\beta\pi\sqrt{C(0)}\lambda
%\end{equation}
%as a rescaling convenient for all three universality classes.

It seems natural to review now known  properties of velocity correlation function. For the reasons that become obvious later it is more convenient to consider first the second derivatives of energy levels
with respect to the parameter, the so called curvatures.

\section{Curvature distributions}
\label{sec:curv}
The curvatures of levels $K_n=\dot{p}_n$ as derivatives of velocities should be called in the dynamics language ``level accelerations''. We stick to curvatures as this is a commonly used concept. Large curvatures appear in the vicinities of avoided crossings in the system. Then, essentially only two levels are involved. Following this strategy \cite{Gaspard90} showed that the  large curvature tail behaves as $|K|^{-(\beta+2)}$ for all three universality classes.
 
The full analysis of curvature distributions, not limited to large curvature tail, was carried out in \cite{Zakrzewski93}. Large numerical data collected for all three ensembles suggested the following simple and analytic form:
\begin{equation} 
P(K)={\cal N}_\beta
 {1\over{\left ( 1+(K/\gamma_\beta)^2\right )^{{\beta+2}\over 2}}}
\label{general}                                                 
\end{equation}
(with $\beta=1,2,4$ for GOE, GUE and GSE, respectively) and 
\begin{equation}
\gamma_\beta=\pi{\beta C(0)\Delta}         
\label{gammas}
\end{equation}
where, recall,  $\Delta$ is the mean level spacing (i.e. an inverse of the mean density of states).
Defining the dimensionless curvature, $k$, as
\begin{equation}
k={K\over\gamma_\beta}
\label{dimless}   
\end{equation}
we have explicitely
\begin{eqnarray}
P_O(K)&=&{1\over2} 
{1\over{\left ( 1+k^2\right )^{3/2}}} \label{goet}\\ 
P_U(K)&=&{2\over{\pi}} 
{1\over{\left ( 1+k^2\right )^{2}}} \label{guet}\\ 
P_S(K)&=&{8\over{3\pi}} 
{1\over{\left ( 1+k^2\right )^{3}}}. \label{gset}
\end{eqnarray}   
These expressions, which could be claimed as being determined via Monte-Carlo integration and inspired guess,  were soon proven analytically for all three 
ensembles of gaussian random matrices \cite{vonOppen94,vonOppen95} and by an alternative technique in \cite{Fyodorov95,Fyodorov11}.

Let us remark that the above definition of $k$ differs from the form suggested by the universality rule, \eqref{uni}, $d^2\epsilon/dx^2$ by a multiplicative factor $\pi\beta$ which simplifies \eqref{goet}-\eqref{gset}. 

The distributions  \eqref{goet}-\eqref{gset} appear to work well for circular ensembles as well as some quantally chaotic systems such as kicked tops \cite{Zakrzewski93} or periodic band random matrices in the metalic regime, as shown by Casati and coworkers \cite{Casati94}.
The question remains, however,  to what extent these RMT predictions are universal and to what extend the particular quantally chaotic systems are faithful to them. 
The first aspect was clarified by Li and Robnik \cite{Li96} who pointed out that a nonlinear transformation from $\lambda$ to some other parameter $\mu(\lambda)$ leads to a different curvature distribution as the
transformation between curvatures is nonlinear. It reads \cite{Li96}:
\begin{equation}
k_\mu= k_\lambda - \frac{p_\lambda}{\pi\beta \langle p_\lambda^2\rangle} \frac{\mu''}{\mu'}.
\label{li}
\end{equation}
In the expressions above $k_\mu$ and $k_\lambda$ are normalized curvatures calculated with respect to the corresponding parameters, $p_\lambda$ - the slope and prime denotes derivative with respect to $\lambda$. As Li and Robnik \cite{Li96} point out since velocities are Gaussian distributed (fast decaying) the universality of curvatures may be restored for large curvatures but nonlinear transformation \eqref{li} prevents universality at all scales, see also \cite{Leboeuf99}. The same argument shows, however, that for any ``local'' linear transformation the universality may hold. As long as changes of $H(\lambda)$ are linear in $\lambda$, as assumed in the derivation above, one might expect the universality to hold.

There is, however, another origin for the lack of universality which gives us insight into the physics involved. 
Already Takami and Hasegawa \cite{Takami92} suggested that the presence of scarring,
i.e. strong localization of eigenstates in the space where unstable periodic orbits exist in the classical limit \cite{Heller84,Bogomolny88} may affect curvatures.  Numerical studies of several examples such as the hyndrogen atom in a magnetic field or quantum billiard proved that this is indeed the case \cite{Zakrzewski93}. While referring the reader to an original paper for numerical details it suffices to say that strong scarring
leads to a pecular level dynamics with some levels (scarred eigenstates) have quite different slope than the rest and interact with other levels only locally in narrow avoided crossings. Those levels behave like solitons and may be described as such \cite{Gaspard89,Nakamura93}.  Their behavior leads to excess of small curvatures (outside of these avoided crossings) as well as very large curvatures (at the centers of avoided crossings).

It is worth stressing (which we just do with a single sentence) that curvatures are strongly linked with transport and conductance  \cite{Casati94,Braun94}. Particularly interesting for this case are situations where the parameter breaks time reversal invariance as it happens for twisted boundary conditions. 

\section{Velocity correlations}
\label{sec:vel}

Let us come back to the level slopes, i.e. velocity correlations.
 Simons and Altschuler  \cite{Simons93,Simons93a} in their analysis considered the autocorrelator of velocities at a some energy difference, $\omega$, $\tilde c(x,\omega)$  (note - this is a different object than $C(x)$ which involves correlations for the same level $n$) that involves all level velocities in a given interval studied. We refer the reader to original papers for details \cite{Simons93,Simons93a}. The $C(\lambda)$ or rather $C(x)$
 was studied numerically  \cite{Zakrzewski95z} for all three unitary classes. A simple analytic approximation for $C(x)$ was proposed in terms of the plasma error function, see    \cite{Zakrzewski95z}.

The large $x$ limits was elegantly solved \cite{Simons93,Simons93a} yielding $C(x)= - 2/\beta \pi^2 x^2$ for GUE.
%with $C_\beta$ being a constant depending on the universality class.
 Interesting information may be obtained 
from a small $x$ limit when   $\tilde c(x,\omega)$ (for $\omega=0$) and $C(x)$ behave similarly. 

Application of Taylor series expansion of $C(x)$ allows one to link the velocity correlator to the variance of the rescales curvatures. Explicitly, one obtains \cite{Zakrzewski95z}
\begin{equation}
C(x) =C(0)(1- \frac{1}{2}\beta^2\pi^2x^2\langle k^2 \rangle).
\end{equation}
Defining normalized correlation $c(x)=C(x)/C(0)$ one
  reproduces the results \cite{Simons93a} for GUE: $c_{\textrm{GUE}}(x)=1-2\pi^2x^2$ and gets $c_{\textrm{GSE}}(x)= 1-\frac{8}{3}\pi^2x^2+ ... $ for the symplectic ensemble \cite{Zakrzewski95z}. Interestingly, for the most common orthogonal universality class one encounters the problem as the variance of curvatures, following \eqref{goet}, does not exist. This indicates that the small $x$ behavior may be singular and the Taylor expansion questionable.
 
 This issue has been studied further in \cite{Guarneri95} where it was shown that in fact $c(x)$ reveals singularities around $x=0$. Taking the parametric dependence \eqref{eq:ham} one may show that Fourier components of the Fourier transform of $c(x)$ has algebraic tails which directly indicates singularities at $x=0$ of the velocity correlator. Again we just quote the the final result which shows that 
 \begin{align}
 c_{\textrm{GOE}}(x) \sim 1+ b_1 x^2|\ln(x)|, \nonumber \\
 c_{\textrm{GUE}}(x) \sim 1-  2\pi^2x^2+ b_2|x^3|,  \\
c_{\textrm{GSE}}(x) \sim 1-\frac{8}{3}\pi^2 x^2+ b_3 x^4 + b_5 |x^5|, \nonumber 
\end{align}
with $b_i$ being coefficients of the order of unity. One may observe that the singularity at $x=0$ becomes weaker with growing level repulsion $\beta$, being most severe for GOE.  An even more in depth analysis of singularities appears in \cite{Walker96} where explicit values for the parameters, $b_i$ are found.

\section{Avoided crossings distributions}
\label{sec:avo}

Another statistical property with interesting links to level spacings is the distribution of avoided crossings, i.e. minima of distances between neighboring levels. The problem of finding the corresponding distribution was formulated by Wilkinson \cite{Wilkinson89} who has shown that integrated distributions for small minimal distances $D$ for GOE (GUE) show similar repulsion as present in spacing distributions. Avoided crossings for billiard models were numerically studied by Goldberg and Schweizer \cite{Goldberg91}. While in a general case the exact distributions are not known in some analytic form, a well working approximations based on 2-levels approximation may be easily derived following the Wigner approach for level spacings themselves \cite{Zakrzewski91}. For GOE case it is 
written down immediately as the 2-level Hamiltonian $H=H_0+\lambda V$ may be written in the eigenbasis of $V$ as
\begin{equation}
 H=\begin{bmatrix}  a & d \\ d & b \end{bmatrix} + \lambda \begin{bmatrix} v_1 & 0 \\ 0 & v_2 \end{bmatrix}.
 \label{eqw}
 \end{equation}
The minimal distance between levels is simply $2|d|$. Since $H_0$ is assumed to correspond to GOE, $d$ is gaussian distributed, so we get the distribution (for $D=2|d|$) normalized to unit mean avoided crossing:
\begin{equation}
P(D)= \frac{2}{\pi}\exp \left[ - \frac{D^2}{\pi} \right ], \ D>0.
\end{equation}
Situation is only slightly more complicated for other ensembles. For GUE $d$ in \eqref{eqw} should be complex, $d=d_1+id_2$,  with independently gaussian distributed  (with the same variance) $d_i$. A simple integral leads to a normalized distribution
\begin{equation}
P(D)=\frac{\pi D}{2} \exp\left [ -\frac{\pi}{4}D^2\right ],
\end{equation}
which is identical to the so called Wigner surmise for spacings for GOE. We observe a simple rule that the avoided crossings in 2-level approximation share the same distribution as the nearest neighbor spacings but the the repulsion parameter $\beta$ reduced by unity. So for GUE with $\beta=2$ we get the Wigner formula corresponding to spacings for $\beta=1$. This is in full agreement with small $D$ perturbative prediction of \cite{Wilkinson89}. Similarly, an explicit calculus shows that for $\beta=4$ GSE the avoided crossing distribution behaves as $D^3$ for small $D$.

Numerical tests (which have to be carefully done to correctly estimate avided crossing values \cite{Zakrzewski93b}) show excellent agreement between 2-level approximate formulae and numerical data for
random matrices of larger sizes. The agreement is in fact better than for the spacings and the Wigner surmize. The reason is simple, the 2-level approximation works better for minimal distances between levels.

.
 
\section{Experiments}

The predictions concerning level dynamics were soon tested, to some extent, in experiments. As typical for quantum chaos, those experiments were not carried out on eigenvalues of the Schr\"odinger
equation but rather on related models of quasi 2D microwave cavities or propagation of acoustic waves. We provide just an uncomplete list of references to beautiful experiments \cite{Stockmann97,Bertelsen99,Dietz06,Poli09,Hul09,Hul09p,Lawniczak13} stressing \cite{Stockmann97} where a rather complete comparison of different measures discussed above with experimental microwave resonance data were carried out. Still one must say that experimental verification of different theoretical predictions is, in this area of physics, quite limited, arguably due to difficulties of collecting sufficient accuracy data for a reliable statistical analysis.

\section{Fidelity susceptibility}
\label{sec:fid}
Rapidly developing in recent twenty years quantum information brought yet another measure which may be related to to level dynamics, the fidelity, ${\cal F}$ \cite{Uhlmann76}. While generally defined for mixed states, for our purposes a pure state definition \cite{Zanardi06}
\begin{equation}
{\cal F} = |\langle \psi(0) | \psi(\lambda)\rangle|
\label{eq:fid}
\end{equation}
where $\lambda$ is the parameter changed. Taylor expansion for small $\lambda$ leads to the definition of fidelity susceptibility, $\chi$
 \begin{equation}
{\cal F}(\hat \rho (0) , \hat \rho(\lambda))= 1 - \frac{1}{2} \chi \lambda^{2} + O(\lambda^{3}).
\label{eq:rozwiniecie}
\end{equation}
Fidelity susceptibility became an indicator of quantum phase transitions.  At the 
transition point,  the ground state properties change  leading to the enhancement of
$\chi$ \cite{Zanardi06,You07,Invernizzi08}.
Apart from ground states, thermal states were also considered
 \cite{Zanardi07,Sirker10,Rams18}). The full fidelity statistics was discussed for the first time in  \cite{Sierant19f}, note that an attempt to identify many-body localization transition is due to \cite{Hu16}.
 We shall review in short the results of \cite{Sierant19f} that provide one of the rare situations when exact analytic results are available for arbitrary size of random matrices. 

Consider $H=H_0+\lambda V$ with both $H_0$, $V$ belonging to GOE or GUE. Fidelity susceptibility of state $|n\rangle$ with energy $E_n$ of $H_0$ is easily derived as
\begin{equation}\label{fidelity}
\chi_n=\sum_{m\ne n}\frac{|V_{nm}|^2}{(E_n-E_m)^2},
\end{equation}	
showing some similarity to curvatures $\dot p_n$ \eqref{e4} - the difference is just a power in the denominator.

The probability distribution of the fidelity susceptibility at energy $E$ reads:
\begin{equation}\label{probability}
P(\chi,E)=\frac{1}{N\rho(E)}\left\langle \sum_{n=1}^N\delta(\chi-\chi_n)\delta(E-E_n), \right\rangle
\end{equation}
which we consider at the center of the spectrum $E=0$.
Following the technique developed in \cite{Fyodorov95,fyodorov12,fyodorov15} one arrives  \cite{Sierant19f} at for GOE case
\begin{eqnarray}
 \nonumber 
 P^O_N(\chi) = 
 \frac{C^{O}_N}{\sqrt{\chi}} \left(\frac{\chi}{1+\chi}
 \right)^{\frac{N-2}{2}}\left( \frac{1}{1+2\chi}\right)^{\frac{1}{2}}\\ 
  \left[ \frac{1}{1+2\chi} + 
 \frac{1}{2} \left(\frac{1}{1+\chi}\right)^{2} 
\mathcal{I}^{O,2}_{N-2}\right],
  \label{eq:5mt}
\end{eqnarray}
where $C^{O}_N$ is a normalization constant and 
\begin{equation}
 \mathcal{I}^{O,2}_{N}=
\begin{cases}
N\frac{N+2}{ N+3/2}, \quad  \quad N \quad \mathrm{even}, \\ 
N+1/2,\quad  \quad N \quad \mathrm{odd}.
\end{cases}
 \label{c9}
\end{equation}
Equation \eqref{eq:5mt} is an exact result for an arbitrary rank $N$ of the random matrix from GOE. This is one of rare situations when analytic formulae for arbitrary $N$ and not only for $N=2$ or $N\rightarrow\infty$ limit are known.

The $N\rightarrow \infty$ limit is interesting. As $ \mathcal{I}^{O,2}_{N}$ is proportional to $N$ one can define  a scaled fidelify susceptibility  $y=\chi/N$.
Its distribution, in the $N\rightarrow\infty$ limit takes a rather simple form
\begin{equation}
 P^O(y)=\frac{1}{6} \frac{1}{y^2}\left(1+\frac{1}{y}\right) \exp\left(-\frac{1}{2y}\right).
 \label{eq: 7mt}
\end{equation}
As tested numerically this expression woks well for $N\sim200$ already.

Similarly, see \cite{Sierant19f} for the derivation, one obtains an analytic, valid for arbitrary $N$ results, for GUE. We quote here just the $N\rightarrow\infty$ limit
for the scaled fidelity susceptibility
 \begin{equation}
 P^U(y)=\frac{1}{3\sqrt{\pi}}\frac{1}{y^{5/2}}\left(\frac{3}{4}+\frac{1}{y} 
 +\frac{1}{y^2}\right) \exp\left(-\frac{1}{y}\right)
 \label{eq: 11mt}
\end{equation}
while for a full expression for $\chi$ valid for arbitrary $N$ as well as for comparison with numerical data we refer to \cite{Sierant19f}.

\section{Generalization to more parameters}
\label{sec:many}

A natural extension of parametric level dynamics occurs in the presence of more than one parameters. One may define
\begin{equation}
H=H_0+ \sum_i \lambda_i V_i
\label{hamnew}
\end{equation}
with $H_0$, $V_i$ statistically independent and drawn from the same (as in this review) or different universality classes. Obviously novel problems to solve appear which we mention briefly only.
Probably it was Michael Wilkinson and his student, Elisabeth Austin, who addressed first 
such a situation in their study of  density of degeneracies, e.g., diabolical points \cite{Wilkinson93}.  Three parameters family
was consider for Chern integer fluctuations \cite{Walker95,Walker96}. Steuwer and Simons \cite{Steuwer98} found the distribution of adiabatic curvature (related to Berry phase) for GUE. Tne multiparameter dynamics was recently revisited by sir Michael Berry and Pragya Shukla who discussed Berry curvature deriving 2-level and 3-level distributions 
\cite{Berry18,Berry20,Berry20a}. They found the large curvature scaling, $P(c)\sim c^{-2}$ for the orthogonal and $P(c)\sim c^{-5/2}$ for the unitary class, the result already present in \cite{Steuwer98}.

 Importantly, however, Berry and Shukla \cite{Berry20a} linked the problem with quantum infomation concept of geometric tensor and the distance between quantum states \cite{Provost80,Venuti07,Kolodrubetz17}. 
The Fubini-Study distance in the Hilbert  space between states differing by a small change of parameters  form $\vec\lambda\equiv (\lambda_1, \lambda_2,..., \lambda_n)$ to $\vec{\lambda}+d\vec{\lambda}$
is 
\begin{equation}
ds^2=1-|\langle n(\vec{\lambda})|n(\vec{\lambda}+d\vec{\lambda})\rangle|^2=\sum_{ij}\textrm{Re}g_{ij}^{(n)}(\vec{\lambda})d\lambda_id\lambda_j,
\label{gt1}
\end{equation}
where $g_{ij}^{(n)}(\vec{\lambda})$ is the so called geometric tensor \cite{Provost80,Venuti07,Kolodrubetz17} which governs the quenches (in $\lambda$) of the system. For \eqref{hamnew}
\begin{equation}
g_{ij}^{(n)}=\sum_{m(\ne n)}\frac{\langle n|V_i|m\rangle\langle m |V_j|n\rangle}{(E_n-E_m)^2}.
\label{gt2}
\end{equation}
Note that the distance between states is determined by the real part of the geometric tensor only - \eqref{gt1}.  The imaginary part, $\textrm{Im}g_{ij}^{(n)}$,  gives the Berry curvature \cite{Berry20a,Penner21}
related to changes of two parameters $\lambda_i,\lambda_j$. For a single parameter problem the geometric tensor reduces to a scalar proportional to the fidelity susceptibility discussed in the previous Section. Also, if $V_i$ belong to same ensembles then the distributions of $g_{ii}$ are equal. One may consider, therefore, the trace $G=\textrm{Tr} g_{ij}^{(n)}$ as an equivalent of  the fidelity susceptibility while the distribution of the imaginary part of the geometric tensor reduces to the Berry curvature distribution.

The distribution of trace, $G$, is, therefore, given by formulae discussed above, valid for arbitrary $N$ \cite{Sierant19f}. The alternative derivation using supersymmetric techniques is provided in \cite{Penner21}
for GUE in the $N\rightarrow\infty$ limit.  The samer authors obtained for the Berry curvature the result derived earlier by \cite{Steuwer98}.

\section{Towards the localization limit}
\label{sec:loc}
While in the introductory part we have considered the transition between iintegrable (localized) regime  in the context of level spacings and the dynamics of Pechukas-Yukawa gas, later we have mainly described the results for the ergodic regime well simulated by Gaussian random ensembles.  Here we shall briefly mention some of the results for level dynamics measures that involve the transition. 

Here the leading at a time contributions were provided by the Como group centered around Giulio Casati \cite{Casati90,Casati91,Casati93,Casati94}. The banded random matrix ensemble provided a natural tool  to stady the transition from ergodic(metallic) to localized transition by varying the width of the band. The team addressed also curvature distributions \cite{Casati94, Guarneri95}. The analytic approach to the problem was pursued by Yan Fyodorov, who, starting around 1994, considered comprehensively level dynamics features close to the localization transition studying velocity correlations \cite{Fyodorov94, Fyodorov95v} or curvature distributions \cite{Titov97}. In particular the velocity distribution for one-dimensional disordered wire is derived using the supersymmetric approach to be
\begin{equation}
P(v_s)=\frac{\pi}{\sinh^2(\pi v_s)}\left\{ \pi v_s \coth(\pi v_s) -1\right\},
\end{equation}
for the scale velocity $v_s$.  The curvature distributions   were more recently addressed
in the context of  MBL studies \cite{Filippone16, Monthus17}. The level dynamics across the many-body localization transition  for the paradygmatic XXZ spin model was considered in \cite{Maksymov19}. 
Velocity, curvatures and fidelity susceptibility distributions were considered. Interestingly while velocities depended on the choice of the parameter (being it the interaction strength or the kinetic tunneling)
curvature exhibited universal behavior in the delocalized regime. In the localized regime curvature distributions reveal system specific characteristics that exemplify the presence of local integrals of motion in the localized phase. Large curvature or large fidelity susceptibility tails change their slope when entering the localized regime. Such a behavior is well understood qualitatively and linked to weaker level repulsion.

\section{Conclusions - where do we stand}
With this travel through last 40 years, starting with Pechukas model \cite{Pechukas83}, we hope to have shown  that many fascinating, breath-taking results have been obtained in the studies of level dynamics of 
complex systems but still there are many open questions and unsolved problems.  There are at least two areas that await a more decisive attack and, hopefully, solutions. One is the transition between different ensembles \cite{Kunstman97}. We have not, on purpose, reviewed few works in this domain for the reader to formulate his own problems. A second related area with several white spots lies in multiparameter level dynamics. The latter has been mostly limited to studies within unitary ensemble, notoriously easiest to treat. We do not know Berry curvature distribution for the orthogonal ensemble. We do not know about the geometric tensor properties when different parameters induce different transitions. Even the simplest questions remain unanswered. For example, simple analysis shows that fidelity susceptibility of Berry curvature decay with reverse quadratic power 
for GOE while the corresponding power is -5/2 for GUE. Can we say something when we couple GOE and GUE ensembles by some parameter, e.g. weakly breaking the time reversal invariance? Can we generalize
the findings to the symplectic ensemble? What about the ten-fold way \cite{Altland97}? It is my believe that we may expect in the future some very interesting results coming from new people entering the subject.
I am already anticipating the excitement.
  
\begin{acknowledgments}
I would like to thank several of the colleagues with whom some of the results presented were obtained. Special thanks are due to Dominique Delande, the work on curvature \cite{Zakrzewski93} and avoided crossing distributions \cite{Zakrzewski93b} done during my stay in Paris formed just the beginning of many years of collaboration with about 50 papers. I profited a lot from discussions with Marek Ku\'s with whom we first treated avoided crossing distributions using the 2-level approach  \cite{Zakrzewski91} and then considered fidelity susceptibility recently \cite{Sierant19f}.  Parametric correlations of velocities were studied in collaboration with Italo Guarneri and Luca Molinari under the guidance of Giulio Casati \cite{Guarneri95} with contributions of Karol \.Zyczkowski who, at that period, collaborated with the Como gang. 
 I want also to mention the recent collaboration with Artur Maksymov and Bitan De.  Last but not least I would like to thank Piotr Sierant, my former PhD student for his contribution to fidelity susceptibility distribution derivation \cite{Sierant19f} and many discussions. %It is the biggest priviledge to observe the development of a student into a world class scientist. 
 Without him this work could not be completed. I am also grateful to  Eugene Bogomolny and Yan Fyodorov for suggestions on the literature of the subject.
The support of  PL-Grid Infrastructure is acknowledged.
This research has been funded by 
 National Science Centre (Poland) under project  2019/35/B/ST2/00034. The support by a grant from the Priority Research Area DigiWorld under the Strategic Programme Excellence Initiative at Jagiellonian University is also acknowledged.

  \end{acknowledgments}

%\bibliographystyle{apsrev4-1}
%\bibliography{ref_21.bib} 

%\include{supplement}
\end{document}